\documentstyle[12pt]{article}
\catcode`@=11
\def\@cite#1{$^{\scriptsize{#1}}$}
\def\@refe#1{#1}
\def\@biblabel#1{{\normalsize\bf{#1}}}
\def\refe{\@ifnextchar
[{\@tempswatrue\@citexr}{\@tempswafalse\@citexr[]}}
\def\@citexr[#1]#2{\if@filesw\immediate\write\@auxout{\string\citation
{#2}}\fi
  \def\@citea{}\@refe{\@for\@citeb:=#2\do
    {\@citea\def\@citea{,}\@ifundefined
       {b@\@citeb}{{\bf ?}\@warning
       {Citation `\@citeb' on page \thepage \space undefined}}%
\hbox{\csname b@\@citeb\endcsname}}}{#1}}
 \catcode`@=12
\def\<{\langle}
\def\>{\rangle}

\title{Combinatorial and topological phase structure of \\
 non-perturbative n-dimensional quantum gravity}
\author{M.Carfora*, M.Martellini**, A.Marzuoli*\\
Dipartimento di Fisica Nucleare e Teorica dell'Universit\`a di
Pavia\\
Via Bassi 6, I-27100 Pavia, Italy* \\
Istituto Nazionale di Fisica Nucleare, Sezione di Roma I,Italy,\\
and Dipartimento di Fisica dell' Universit\'a di Milano\\
Via Celoria 16, I-20133 Milano, Italy (permanent address) ** \\
Istituto Nazionale di Fisica Nucleare, Sezione di Pavia,Italy*,**}
\def\Ricco{{\cal {R}}(n,r,D,V)} %this is a macro for a formula%
\begin{document}
\maketitle
\begin{abstract}
 We provide a non-perturbative geometrical characterization of
the partition function of $n$-dimensional  quantum gravity based
on a coarse classification of  riemannian geometries. We show
that, under natural geometrical constraints, the theory admits a
continuum limit with a non-trivial phase structure parametrized
by the homotopy types of the class of manifolds considered. The
results obtained qualitatively coincide, when specialized to
dimension two, with those of two-dimensional quantum gravity
models based on random triangulations of surfaces.
\end{abstract}
\vfill\eject

\section{Introduction}
      In this paper  we present  results which extend to any
dimension $n \geq  2$ some of the non-
perturbative aspects of two-dimensional quantum gravity theories
based on random triangulations of surfaces\cite{Brez,Doug,Gros}.\par
This extension relies on previous work of two of us on functional
integration on the space of riemannian structures\cite{Carf,Marz},
and it is made possible  by exploiting the
properties of the  space  of  bounded  geometries  introduced  by
M.Gromov \cite{Grom}. In particular, we shall make use of
 the infinite dimensional set, $\Ricco$,  of  all  compact  riemannian
manifolds  of dimension $n$  (not necessarily of the same underlying
topology), with sectional curvature bounded below by $-(n-1)r$,
diameter
bounded
above by $D$  and volume  bounded below  by $V$  ($r$  being  any
real
number, and  $D$, $V$ any positive real numbers)\cite{Grom}.
 $\Ricco$ is
an  infinite-dimensional   space which  can be  endowed with a
natural  topology   most  suitable  for  handling  questions  of
convergence of  random triangulations  on  riemannian  manifolds.
A property, this latter, connected to the observation that  for
any manifold $M$ in such a  class it is possible
to  introduce  coverings   by  geodesic   balls  whose
combinatorial patterns  yields simplicial  approximations  which
provide a  coarse classification of the riemannian   structures
occurring
 in $\Ricco$\cite{Grov}.\par
\vskip 0.5 cm
In what follows we need  few details  on such  coverings, thus,  let
$L(m)
\equiv1/m$, denote  a constant  (later on to  be interpreted  as
a cut-off related to the spacing of a triangulation). For any given
$L(m)$,
and any  given manifold $M$ in $\Ricco$, we  can  always introduce
on $M$  geodesic balls  coverings in  such a  way that  a  finite
collection of open geodesic
balls of radius $2L(m)$ covers $M$, while the corresponding balls
of radius $L(m)$ are disjoint, (this particular covering is often
called a minimal $L(m)$-net for $M$). Two combinatorial
invariants characterize  such construction. The first is the
{\it filling function} $N_{(m)}(M)$  of the  covering, {\it  i.e.},
the
function which associates  with $M$  the maximum  number of
geodesic balls realizing a minimal $L(m)$-net on $M$.\par
The second invariant is  the (first) {\it intersection  pattern} of
the
covering,
${\gamma}_{(m)}(M)$, defined by  the collection  of pairs  of geodesic
balls of radius $2L(m)$ having  a non  empty  pairwise  intersection.
These  two
invariants  characterize     the   $1$-skeleton,
${\Gamma}_{(m)}(M)$, of the geodesic  balls covering. Similarly, upon
considering   the higher  order intersection patterns,  one ends
up  in  defining  the  two-skeleton $K_2(M)$, and eventually  the
{\it nerve} of the covering of   the  manifold   $M$. If  $m$ is
sufficiently  large this nerve gives rise to a
polytope  which can be assimilated to a symplicial approximation of
the  manifold  $M$.  In  particular,  the $1$-
skeleton is just the vertex-edge structure  of this
approximation.\par
\vskip 0.5 cm
Any two  manifolds $M_1$ and   $M_2$  in  $\Ricco$ endowed  with
minimal  $L(m)$-net are considered equivalent if and only if they
have  the  same  filling  functions  and  the  same  intersection
patterns up to combinatorial isomorphisms.  Such   relation
partitions   $\Ricco$  into   disjoint
equivalence classes  whose number  can be  shown  to  be {\it
finite}, function  of   the  parameters   $m$,  $n$,$r$,  $D$,
$V$ \cite{Grov}.  Each equivalence  class of manifolds is
characterized by the abstract (unlabelled)    graph  ${\Gamma}_{(m)}$
defined  by the  $1$-skeleton of the $L(m)$-covering. The  order of
any
 such graph ({\it i.e.}, the number of
vertices) is  provided by  the filling  function $N_{(m)}$, while
the structure  of the  edge set  of ${\Gamma}_{(m)}$  is defined
by  the intersection pattern ${\gamma}_{(m)}(M)$. It is important
to remark that  on $\Ricco$ either the filling function or the
intersection pattern cannot  be arbitrary. The former is always
bounded
 above for each  given $m$, (for instance, if $r<0$ an upper
bound is provided by $D^nm^n$),and  the best  filling of
a riemannian manifolds  with geodesic  balls  of  radius  $1/m$
is realized on (portions of)  spaces  of  constant  curvature
\cite{Grom}.  The latter is  similarly  controlled  through  the
geometry
 of the  manifold to  the effect  that the {\it average degree},
$d(\Gamma)$,  of  the  graph ${\Gamma}_{(m)}$, ({\it
i.e.}, the  average number of edges incident on a
vertex of  the graph), approaches  a constant value  as the
radius of the balls defining  the covering  tend to  zero, ({\it
i.e.},
as $m \to \infty$).  Such constant  is independent  from $m$,
and can be estimated in terms of the parameters $n$, $r$, $D$,
and $V$ \cite{Grov}. \par
An intuitive idea of the natural topology of the Gromov
space of bounded geometries  can  be grasped by
noticing that two riemannian manifolds in $\Ricco$ get closer and
closer in  such topology  if we  can introduce  on them finer and
finer minimal $L(m)$-nets of geodesic balls with the same filling
function and  the same  intersection pattern,  {\it i.e.}, if the
symplicial approximations induced by geodesic balls coverings have a
similar
vertex-edge structure. \par
\vskip 0.5 cm
Within this  geometrical framework, let us consider the
following graph-theoretical statistical sum
\begin{eqnarray}
{\Xi}(m,z)  =   \sum_{\Gamma}z^{N_{(m)}(\Gamma)}   \exp
{[\beta{\Phi}_{(m)}(\Gamma)]} \label{uno}
\end{eqnarray}
\noindent   where $z  \equiv \exp (-c)$, and
$\beta$ are constants,
and where the (finite) sum  is over  all inequivalent  $1$-
skeletons ${\Gamma}_{(m)}$, with order     $N_{(m)}(\Gamma)$ and
size ${\Phi}_{(m)}(\Gamma)$, realized by the  possible $L(m)$-
geodesic balls covering  over manifolds  in $\Ricco$,
(the size ${\Phi}_{(m)}(\Gamma)$ is the number of edges in the
graph ${\Gamma}_{(m)}$). Notice that either the filling
function $N_{(m)}$ or the size ${\Phi}_{(m)}(\Gamma)$, besides
 having a natural combinatorial
meaning for the manifolds in $\Ricco$, can be directly related to
 riemannian invariants. A standard
expansion for the riemannian volume of small geodesic balls
yields $N_{(m)}(M) \simeq B(n,m)\{1+[m^{-2}/6(n+2)]R+o(m^{-
2})\}$, where $B(n,m)$ is a constant depending on the volume of
$M$, $R$ is the scalar curvature evaluated at a suitable point
and $o(m^{-2})$ stands for
higher order terms which are universal polynomials in the
curvature tensor and its covariant derivatives. A similar
expansion holds true also for ${\Phi}_{(m)}(\Gamma)$. In
particular, from the properties of geodesic balls coverings
it follows that the size of the graph $\Gamma$ grows with $m$
as $N_{(m)}$ does, and for $m$ sufficiently large one can
write
${\Phi}_{(m)}({\Gamma}(M)) \simeq B_{M,R}N_{(m)}({\Gamma}(M))$,
where $B_{M,R}$ is a constant, independent from $m$, related to the
curvature of $M$. Collecting such results, we can heuristically
rewrite ${\Xi}(m,z)$, to leading order in $m$, as
\begin{eqnarray}
{\Xi}(m,z)\simeq \sum_{\Gamma}\exp[m^n(-c\cdot {Volume}
+\beta\cdot {Curvature})]\nonumber
\end{eqnarray}
Since the above summation over all graphs $\Gamma$ is a sum over all
finite equivalence classes of manifolds  in $\Ricco$  parametrized
by their one-skeletons, we can regard ${\Xi}(m,z)$ as a discrete
version
of a sum over manifolds in $\Ricco$ with a combinatorial weight
related to volume and curvature. \par
\vskip 0.5 cm
\section{Geodesic balls coverings and lattice gas
statistical mechanics}
{}From a thermodynamical point of view,  ${\Xi}(m,z)$ has the
structure of  a grand-partition  function computed  for a lattice
gas  with   negative  pair   interactions  evaluated  at inverse
temperature $\beta$  and for  a fugacity $z$.\par
In order to discuss this statistical equivalence more in details
we explicitly identify the geodesic balls of a minimal $L(m)$-net on a
manifold  $M$ of bounded geometry with the vertices
 $\{p_1^{(0)},\ldots p_N^{(0)}\}$,
 of an abstract graph ${\Gamma}(M)$. Such ${\Gamma}(M)$ is
defined by connecting the $\{p_i^{(0)} \}$ among themselves with
undirected
edges $p_{ij}^{(1)}=\{p_i^{(0)}, p_j^{(0)} \}$ if and only if the
geodesic balls
labelled $p_i^{(0)}$ and
$p_j^{(0)}$ have a non-empty intersection when we double their radius.
Any
two such graph are considered equivalent if, up to the labelling
$\{p_i^{(0)} \}$ of the balls, they have the same vertex-edge scheme,
{\it viz.} we are considering unlabelled graphs associated
with the possible one-skeletons of geodesic balls coverings of
manifolds of bounded geometries.
For any given $m$, any such graph can be topologically
imbedded in a larger regular,({\it {i.e.}},
whose vertices have all the same degree), graph
${\Omega}_{(m)}$, ($\Omega$ for short), having as its
connected subgraphs all possible $1$-
skeletons ${\Gamma}_{(m)}$ which are realized by $L(m)$-geodesic
balls coverings as $M$ varies in $\Ricco$. One may think of such
$\Omega$ as the configuration space for the finite set of
possible $1$-skeletons that can
be realized by minimal nets on the manifolds in $\Ricco$.\par
For a given $m$, a graph realizing $\Omega$ can be constructively
defined  first by labelling the graphs ${\Gamma}_{(m)}$ and
order them  by inclusion,
{\it i.e.}, ${\Gamma}^{(i)}_{(m)} \subset {\Gamma}^{(j)}_{(m)}$ if
${\Gamma}^{(i)}_{(m)}$ is a subgraph of ${\Gamma}^{(j)}_{(m)}$,
and then by considering the union $H_{(m)}$ over the
{\it prime} graphs ${\Gamma}^{(k)}_{(m)}$, ({\it i.e.}, over
those graphs which do not appear as subgraphs of other
${\Gamma}_{(m)}$):
\begin{eqnarray}
H_{(m)} \equiv {\cup}^*_j {\Gamma}_{(m)}^{(j)}
\label{acca}
\end{eqnarray}
Let $d({\Omega}_{(m)})$ the maximum degree of the vertices of
$H$, (or which is the same, the maximum degree of the vertices among
those occurring in the various ${\Gamma}_{(m)}^{(j)}$). Notice that,
for every $m$, we have $d(\Omega) \leq C_{n,r,D}$,
where the constant $C_{n,r,D}$ is the upper bound to the average
degree of
the graphs ${\Gamma}_{(m)}$ associated with the
 $L(m)$-geodesic balls realized in $\Ricco$.\par
The existence of
the graph ${\Omega}_{(m)}$ is assured by a theorem of P.Erdos and
P.Kelly\cite{Erdo},
and  in order to construct it,  we proceed as follows\cite{Carf}.
First add
to $H_{(m)}$ a set $I$ of $b_{(m)}$ isolated points,(the number of
which, as indicated, is a fuction of the given $m$), a new graph is
formed from $H_{(m)}$
and $I$ by adding edges between pairs of points in $I$ and $H_{(m)}$,
(notice that no edges are added between vertices in $H_{(m)}$). The
strategy is to give rise, in this way, to a new graph which is
regular of degree $d({\Omega}_{(m)})$ while keeping the
number $b_{(m)}$ of vertices to be added to $H_{(m)}$ as small as
possible.
As expected, the number $b_{(m)}$ of added vertices depends only on
the
degree sequence of the graph $H_{(m)}$. In particular if $d_i$
denote the various degrees at the vertices of $H_{(m)}$,
then $b_{(m)}$ is the least integer satisfying: {\it (i)}
$b_{(m)}d({\Omega}_{(m)})\geq \sum_i(d(\Omega)-d_i)$, {\it (ii)}
$b_{(m)}^2-(d(\Omega)+1)b_{(m)}+\sum_i(d(\Omega)-d_i) \geq 0$, {\it
(iii)}
$b_{(m)} \geq \max (d(\Omega)-d_i)$, and {\it (iv)}
$(b_{(m)}+h_{(m)})d(\Omega)$ is even, where $h_{(m)}$ is the order
of the union graph $H_{(m)}$\cite{Erdo}.\par
It should be noted that in the large $m$ limit, ${\Omega}_{(m)}$ is
a regular graph whose degree,
$d({\Omega}_{(m)})$, is  independent
from $m$, being bounded above in terms of $n$,$r$, $D$, by
$(C_{n,r,D})^{n-1}$.\par
\vskip 0.5 cm
With these preliminary remarks we can work out the correspondence
between minimal geodesic balls coverings of manifolds  $M$ in
$\Ricco$ and the statistical system described by a lattice
gas on ${\Omega}_{(m)}$ by identifying
 the collection of
 geodesic balls,   providing the dense packing of $M$, with a gas  of
indistinguishable particles.
 Each particle can
  occupy at most one  site of  ${\Omega}_{(m)}$,
    (the geodesic balls of radius $L(m)=1/m$ are
  disjoint), and  the configuration of occupied sites corresponding to
the net of
  balls covering $M$ is  defined by the vertices of the graph
  ${\Gamma}(M) \subset {\Omega}_{(m)}$.  \par
    The interaction energy between two occupied sites is
supposed
  to be different from zero only if the sites in question  are
connected by an edge $\{p_i^{(0)}, p_j^{(0)}\}$  of ${\Gamma}(M)$,
 namely  if  the points of
the minimal $L(m)$-net
corresponding to the  lattice sites in question are in the
intersection pattern
 of the manifold  according to the definition recalled above, (the
 geodesic balls $p_i^{(0)}$ and $p_j^{(0)}$ have a non-empty
 intersection when their radius is doubled). More in general,
 $p_i^{(0)}$ and $p_j^{(0)}$ will be said to be
neighbors if $(p_i^{(0)},p_j^{(0)})$ is an edge of the graph
${\Omega}_{(m)}$. Obviously, ${\Omega}_{(m)}$ will contain as
possible configurations of occupied sites, ({\it i.e.}, as possible
subgraphs), not only those graphs ${\Gamma}(M)$ representing
one-skeleton of manifolds in $\Ricco$, but also graphs
associated to the $b(m)$ added vertices (and to the corresponding
edges), needed in order to construct ${\Omega}_{(m)}$.
In any case,
given a configuration of occupied sites in ${\Omega}_{(m)}$
represented
by a graph
 ${\Gamma}$ with $N_{(m)}(\Gamma)$ vertices,
  the total interaction energy corresponding to such
configuration is  given   by
\begin{eqnarray}
E({\Gamma}) =  - e_0{\Phi}_{(m)}({\Gamma})
\end{eqnarray}
\noindent  where $e_0$ is a constant  (which provide the scale of
energy,
and which here and henceforth we set equal to one)
and
${\Phi}_{(m)}$ is the  number of edges in the graph
${\Gamma}$,
{\it viz.}, when dealing with one-skeleton graphs, the total
number of pairs $(i,j)$
belonging to the intersection  pattern of the minimal $L(m)$-net
$\{p_1^{(0)}, \ldots , p_N^{(0)}\}$.
The corresponding Boltzmann weight (as a lattice gas) is
then
\begin{eqnarray}
\exp [- {\beta}E({\Gamma}] = \exp [{\beta}
{\Phi}_{(m)}({\Gamma})]
\end{eqnarray}
Roughly speaking,
according to the correspondence just established the
configurations of the  lattice gas considered  are labelled by random
graphs representing the possible $1$-skeletons  ${\Gamma}_{(m)}$
of manifolds with bounded geometry, and,
for a sufficiently large $m$, we get a  grand-ensemble of
such configurations whose
 thermodynamical parameters keep track of the average volume
 and curvature  of the manifolds in $\Ricco$.\par
\vskip 0.5 cm
It must be stressed that here we are
dealing with  a statistical  system not  evolving  on  a  regular
lattice, for, it shows up more complex interactions which
correspond to the non-trivial topology of the
manifolds underlying  the coverings. For  manifolds  in
$\Ricco$, the boundedness properties for the filling function and
the cardinality of the  intersection   pattern imply that such
interactions  are
localized and do not yield for a catastrophic behavior when going to
the  thermodynamic limit.  From a  more  technical  side,
one proves the existence of such limit by exploiting the
compactness properties of $\Ricco$. These properties easily
allow to show that $\lim_{m \to \infty} {\Xi}(m,z)$ exists even
if in general this limit is not unique.  Such  non-uniqueness
corresponds to  the possible  onset of  phase transitions  in the
system. To be more specific,
let us  take advantage of the introduction of the graph
${\Omega}_{(m)}$
in order to rewrite
${\Xi}(m,z)$ as
the polynomial
\begin{eqnarray}
{\Xi}(m,z) = \sum_{\Gamma}{\hat z}^{N_{(m)}(\Gamma)}\prod_{p \in
\Gamma}\prod_{q \in \Omega \backslash \Gamma}
\exp [- {1 \over 2}\beta A_{\Omega \backslash \Gamma}(p,q)]
\label{Lee-Yang}
\end{eqnarray}
where $A_{\Omega \backslash \Gamma}(p,q)$ is the adjacency matrix
of the graph $\Omega \backslash \Gamma$,
({\it i.e.},the matrix whose entries are
$1$ if the vertices $p$ and $q$ of the graph in question are
connected by an
edge, zero otherwise; with a slight abuse of
notation, we have denoted by $\Omega \backslash \Gamma$
 the graph in $\Omega$ obtained by removing all the edges, but not
the vertices, belonging to $\Gamma$).
 We have also introduced a {\it normalized fugacity} ${\hat z}$
according to
\begin{eqnarray}
{\hat z} \equiv z \exp [{1 \over 2} \beta d(\Omega)]
\label{fugacity}
\end{eqnarray}
(Notice that in general ${\hat z}$ depends from the given $m$
through the expression of the degree $d({\Omega}_{(m)})$, however,
for $m$ sufficiently large this dependence will eventually disappear).
We wish to stress again that the sum (\ref{Lee-Yang}) is {\it
restricted}
only to those graphs ${\Gamma}_{(m)}$ which are $L(m)$-geodesic
balls one-skeletons for manifolds in $\Ricco$, and it is
{\it not extended}
to all possible subgraphs of ${\Omega}_{(m)}$. The {\it unrestricted
sum}, ${{\Xi}(m,{\hat z})}^*$, is  given by
\begin{eqnarray}
{{\Xi}(m,{\hat z})}^* = \sum_{\forall\Gamma \subset \Omega}
{\hat z}^{N_{(m)}(\Gamma)}\prod_{p \in
\Gamma}\prod_{q \in \Omega \backslash \Gamma}
\exp [- {1 \over 2}\beta A_{\Omega \backslash \Gamma}(p,q)]
\label{Lee-Yang2}
\end{eqnarray}
and we can formally write
\begin{eqnarray}
{{\Xi}(m,{\hat z})}^* = {\Xi}(m,{\hat z})+ \sum_{\Omega \backslash
f(\Ricco)}
{\hat z}^{N_{(m)}(\Gamma)}\prod_{p \in
\Gamma}\prod_{q \in \Omega \backslash \Gamma}
\exp [- {1 \over 2}\beta A_{\Omega \backslash \Gamma}(p,q)]
\label{star}
\end{eqnarray}
where the sum over ${\Omega \backslash f(\Ricco)}$ indicates summation
over
all graphs in ${\Omega}_{(m)}$ which are not $L(m)$-one-skeletons for
manifolds in $\Ricco$.\par
Notice that in the construction of ${\Omega}_{(m)}$ out of the
one-skeleton graphs ${\Gamma}_{(m)}(M)$ with the Erdos-Kelly algoritm,
(see (\ref{acca})), no edges are introduce between the graphs
${\Gamma}_{(m)}(M)$, new edges may occurr only between the $b_{(m)}$
added vertices and between these latter vertices and the graphs
${\Gamma}_{(m)}(M)$. From such remarks it follows that the graphs in
${\Omega}_{(m)}\backslash f(\Ricco)$ are defined by
the $b_{(m)}$ added vertices and by the possible edges between them
and the graphs ${\Gamma}_{(m)}$.\par
The relation (\ref{star}) between the unrestricted, ${\Xi}^*$, and the
restricted
statistical sum ${\Xi}$, is useful for suggesting the
boundary conditions which uncover the non-trivial phase structure of
$\lim_{m \to \infty}{\Xi}(m,{\hat z})$.\par
In order to proceed in this direction, let us notice that the
unrestricted statistical sum (\ref{Lee-Yang2}),
${{\Xi}(m,{\hat z})}^*$, takes on the classical polynomial
 Lee-Yang structure\cite{Ruel}.
It follows, according to the Lee-Yang circle theorem \cite{Ruel},
that the zeroes of ${{\Xi}(m,{\hat z})}^*$, thought of as a function
of the complex variable ${\hat z}$, all lie, for each given
value of $m$, on the circle
 $\{{\hat z} \colon \vert {\hat z} \vert =1 \}$ in the plane of the
complexified fugacity
${\hat z}$.\par
\section{Critical behavior}
The Lee-Yang type result quoted above
relates the presence or absence of a phase transition to the analycity
properties of the free energy associated with the grand-partition
function ${\Xi}^*$. A rather delicate analysis\cite{Carf},
shows that it is possible to carry out the thermodynamic
limit on  ${\Xi}^*$, with the boundary condition of keeping empty the
$b_{(m)}$ added sites.
(Geometrically speaking, this boundary condition favors the sampling
of subgraphs of ${\Omega}_{(m)} \uparrow {\Omega}_{\infty}$ which
represents geodesic balls coverings of manifolds in $\Ricco$).
 Then, on applying a variant of the Pirogov-
Sinai theorem \cite{Ruel} on the occurrence of phase transitions
in lattice
gases with attractive pair interactions, one can show that at
sufficiently low temperature the statistical system described
by ${\Xi}$  undergoes a phase transitions, in the large $m$ limit, as
 ${\hat z}\to 1^{-}$.\par
  The interpretation in geometrical terms of the
critical behavior of ${\Xi}(m,{\hat z})$ as $m \to \infty$ and
${\hat z}\to 1^-$ relies on a deep result on the structure of
$\Ricco$. This result \cite{Grov}  states that, for $m$
sufficiently large, manifolds in $\Ricco$ with the same
$1$-skeleton ${\Gamma}_{(m)}$ are homotopically equivalent, and
the number of different homotopy-types of manifolds
realized in $\Ricco$ is finite. In particular, if $m_0$ is the
smallest $m$ such that geodesic balls of radius smaller than
$2L(m_0)$ are contractible, then $\Ricco$ contains less than
$[{\hat N}_{(m_0)}]^4$ distinct homotopy types, where
 ${\hat N}_{(m)}$ denotes an upper bound to the filling function
over $\Ricco$. In connection with this homotopy finiteness
theorem, it can be  explicitly shown\cite{Carf} that the
continuum limit of the sequence of measures associated with
${\Xi}(m,z)$ yields for at least two limit distributions for
sufficiently low temperatures. Such distributions are parametrized
by the fundamental group of the manifolds  sampled in $\Ricco$
by ${\Xi}$.\par
It is possible to provide a less formal interpretation of the
critical behavior of ${\Xi}$ which, even if less rigorous than the
one just sketched, has the advantage of a more direct connection
with the two-dimensional theory.
To this end let us introduce the function $B_{\lambda}({\Gamma}_{(m)},
{\pi}(M))$ which counts the number of combinatorially
inequivalent
$1$-skeletons ${\Gamma}_{(m)}$ with $\lambda$ vertices and with
given average degree $d_{\lambda}$, which can
 be drawn on a manifold $M$ in the homotopy class $\{{\pi}(M)\}$. This
function depends only on the fundamental group ${\pi}_1(M)$ of $M$.
Its
asymptotic behavior, as $\lambda \to \infty$, is well known for
surfaces of given genus \cite{Pari}, and it is one of the main
ingredient for understanding the double-scaling limit in
two-dimensional quantum gravity models. A qualitatively  similar
behavior  can be obtained here too for $B_{\lambda}$, by
rewriting ${\Xi}(m,z)$ as $\sum_{\lambda}{\hat
{z}}^{\lambda}a_{\lambda}$
 with $a_{\lambda}
\equiv
\sum_{\{{\pi}(M)\}}\sum_{d_{\lambda}}B_{\lambda}({\Gamma}_{(m)},
{\pi}(M))\exp
[1/2 \beta(d_{\lambda}- d(\Omega))\lambda]$.

Notice that the summation over the average degrees,
$\sum_{d_{\lambda}}$,
extends, independently of $\lambda$, over a finite number of terms
since
we have the uniform bound $d_{\lambda} \leq C_{n,r,D}$, which does not
depend on $m$,
(the possibility of the above
rewriting
readily
follow from (\ref{uno})  by expressing the sum
over the $1$-
skeleton ${\Gamma}_{(m)}$ as the {\it finite} sum over the
distinct homotopy types $\{{\pi}(M)\}$ sampled in $\Ricco$). By
Lee-Yang theorem, the power series ${\Xi}({\hat z}) \equiv
\lim_{m \to \infty} \sum_{\lambda}{\hat z}^{\lambda}a_{\lambda}$
has radius of convergence one; {\it i.e.}, $\limsup_{\lambda \to
\infty}(a_{\lambda})^{1/ \lambda} =1$, which yields, for each
homotopy type,
\begin{eqnarray}
 \limsup_{ \lambda \to
\infty}\frac
{\ln B_{\lambda}({\Gamma}_{(m)})}{\lambda} =
\limsup_{m \to \infty}
\frac{1}{2} \beta [d({\Omega}_{(m)}) - d_{\lambda}]
\equiv \ln {\Lambda}_c \nonumber
\end{eqnarray}
(for the sake of simplicity, here we are tacitly assuming that for a
given
${\lambda}$, just one average degree $d_{\lambda}$ is dominating in
$a_{\lambda}$. The argument can be extended to the situation where
distinct degrees do contribute. In the two-dimensional theory, this
latter circumstance corresponds to surfaces discretized with different
percentages of $n$-gones, say triangles, squares, and pentagons).\par
The above relation implies that asymptotically in
$\lambda$, $B_{\lambda}$ behaves
as $({\Lambda}_c)^{\lambda}\exp {o(\lambda)}$.
The geometrical meaning of the terms $o(\lambda)$ in this asymptotics
 follows by noticing
that ${\Gamma}_{(m)}$ is a graph drawn on a two-dimensional
complex $K_2(M)$, (the two-skeleton of M), whose fundamental
group is isomorphic to ${\pi}_1(M)$. This implies that the number
of edges of ${\Gamma}_{(m)}(M)$ is given by
${\Phi}_{(m)}({\Gamma}(M))=\lambda + h[{\pi}_1(M)]+ {\rho}(\Gamma) -
1$,
 where  $h[{\pi}_1(M)]$ is the number of non-trivial generators
of ${\pi}_1(M)$, and where ${\rho}(\Gamma)$ is the number of
relations in ${\pi}_1(M)$ obtained by pasting faces
 onto the graph so as to get $K_2(M)$. [To be more precise,
the inclusion ${\Gamma}_{(m)} \to K_2$ induces an epimorphism
between the free fundamental group of the graph ${\Gamma}_{(m)}$
and the fundamental group of $K_2$. This implies that this latter
fundamental group, which is isomorphic to the fundamental
group of the manifold $M$, has a presentation with
$1+{\Phi}_{(m)}(\Gamma)-N_{(m)}(\Gamma)$ generators and
${\rho}(\Gamma)$
relations. It can be shown that $h[{\pi}_1(M)]$ is the (finite) number
of geodesic loops generators, (with given basepoint), non homotopic
to zero, representing the inequivalent homotopy classes of
${\pi}_1(M)$].\par
{}From the properties of the geodesic balls coverings it follows that
$\lim_{\lambda \to \infty}{\Phi}_{(m)}/{\lambda}$ exists and it is
finite. Thus, for large $\lambda$, and for any manifold $M$ in
$\Ricco$, we can write ${\rho}({\Gamma}(M))={\alpha}_h(M)\lambda +
{\sigma}_h(M)$ where we have introduced the constants ${\alpha}_h$ and
${\sigma}_h$ which in general depend on the number of non trivial
 generators $h$
of ${\pi}_1(M)$.According to these remarks we can eventually write
the number of edges of ${\Gamma}(M)$ in terms of the non-trivial
generators of ${\pi}_1$ as
\begin{eqnarray}
{\Phi}_{(m)}({\Gamma}(M)) =(1+{\alpha}_h(M))\lambda +h[{\pi}_1(M)]+
{\sigma}_h(M)-1
\end{eqnarray}
Such an expression shows that with each graph ${\Gamma}$, representing
possible one-skeletons of geodesic balls coverings, we may
formally associate a graph ${\hat {\Gamma}}$ with the same number of
edges but with ${\hat {\lambda}}\equiv (1+{\alpha}_h(M))\lambda$
vertices.
In particular, it follows that the counting function
$B_{\lambda}({\Gamma}_{(m)})$ is proportional to the
number of unlabelled graphs with ${\hat {\lambda}}$ vertices and
${\Phi}_{(m)}$
edges.\par
In order to count such graphs we can use
P\'olya's enumeration theorem \cite{Hara}. A direct computation
allows to  show that the following estimate holds true
\begin{eqnarray}
C_1\{(C_2)^{\lambda}{\lambda}^{h[{\pi}_1(M)]+{\sigma}_h-5/2}\} \leq
B_{\lambda}
\leq C_3 \{(C_4)^{\lambda}{\lambda}^{h[{\pi}_1(M)]+{\sigma}_h-3/2}
+O({\lambda}^{3/2}\exp [{\lambda}^2])\}\nonumber
\end{eqnarray}
\noindent where $C_1$,$C_2$,$C_3$,$C_4$ denote suitable constants.
A comparision of this estimate with the asymptotics for
$B_{\lambda}$, derived from Lee-Yang
theorem,  allows to show that the term $\exp[o(\lambda)]$ in
such asymptotics is given by
 ${\lambda}^{h[{\pi}_1(M)]+{\sigma}_h- \xi}$, where $\xi$ is a
parameter with  $3/2 \leq \xi \leq 5/2$.\par
Since we are counting inequivalent graphs which can be
drawn  on a two-dimensional object , (the two-skeleton $K_2$), the
asymptotics for $B_{\lambda}$, as derived above, is, as expected,
 consistent with the asymptotics that could have
been obtained by graphs counting matrix-models techniques.
More in particular, if we set, (as we do henceforth, for simplicity),
 $\xi = 2 -{\gamma}_0$ and
${\sigma}_h(M)=-(1/2){\gamma}_0h[{\pi}_1(M)]-1$ where ${\gamma}_0$ is
a parameter with $-(1/2)\leq {\gamma}_0\leq (7/2)$, then
$B_{\lambda}\sim {\rho}_h({\Lambda}_c)^{\lambda}
{\lambda}^{(1-h/2)({\gamma}_0-2)-1}$,
(where ${\rho}_h$ is a suitable constant),
 which is the usual asymptotics
for the number of unlabelled graphs (with $\lambda$ vertices) that can
be
drawn on a two-dimensional surface of genus $(1/2)h[{\pi}_1(M)]$.\par
According to such asymptotic behavior for $B_{\lambda}$,
a Tauberian theorem,  (actually an Abelian regularity
results),  yields for ${\Xi}({\hat z})$, as ${\hat z}
\to 1^-$, the expression
\begin{eqnarray}
{\Xi}({\hat z}) \sim \sum_{\{{\pi}_1\}} \sum_{\lambda}
{\rho}_h{\lambda}^{(1-h/2)({\gamma}_0-2)-1}=
\sum_{\{{\pi}_1\}}{\rho}_h{\zeta}[1-(1-h/2)({\gamma}_0-2)] \nonumber
\end{eqnarray}
\noindent  where ${\zeta}(\ldots)$ denotes Riemann's zeta function.
This expression is valid as long as $(1-h/2)({\gamma}_0-2)<0$ for all
homotopy types sampled by ${\Xi}({\hat z})$. If
$(1-h/2)({\gamma}_0-2)>0$, then it follows that ${\Xi}({\hat z})$
behaves critically as
\begin{eqnarray}
{\Xi}({\hat z})\sim \sum_{\{{\pi}_1\}}{\rho}_h
\frac{{\Gamma}[({\gamma}_0-2)(1- h/2)]}
{[c-(1/2){\beta}d(\Omega)]^{({\gamma}_0-2)(1-h/2)}}
\label{Crit}
\end{eqnarray}
when $[c-(1/2){\beta}d(\Omega)] \to 0^+$, (notice that in the above
expression ${\Gamma}(\ldots)$ denotes the Euler Gamma function, and
that
the sum is restricted to those homotopy types  for which the above
positivity condition holds true).\par
 The analysis of these results is
particularly simple and geometrically expressive if  we assume
that $({\gamma}_0-2)<0$, the most general case can be then obtained
with obvious
modifications.\par
Under such hypothesis it immediately follows that if all homotopy
types
sampled in $\Ricco$ by ${\Xi}({\hat z})$ have trivial fundamental
group, {\it i.e.}, if $h[{\pi}_1]=0$, then as ${\hat z} \to 1^-$,
${\Xi}({\hat z})$ converges to
$\sum_{\{{\pi}_1\}} {\zeta}[1-({\gamma}_0-2)]$.
We are
sampling simply-connected manifolds, and we have a pure phase.
We still have regular behavior if the fundamental group
has a non-trivial generator, {\it i.e.}, if $h[{\pi}_1]=1$.
But as
soon as one of the homotopy type sampled by ${\Xi}({\hat z})$ has
two non trivial generators, {\it i.e.}, as $h[{\pi}_1]=2$
for some homotopy type, then as ${\hat z}\to 1^-$ we get a
phase transition associated with the simple pole of the
corresponding ${\zeta}[1-(1-h/2)({\gamma}_0-2)]$.
Roughly speaking we can interpret this phase transition as the
attachement of a handle to the two-skeletons of the
manifolds in the homotopy types being sampled. When the number
of generators of the fundamental group further increases, a
correspondingly increasing number of distinct types of non-simply
connected manifolds contributes to the grand-partition function
and, as ${\hat z}\to 1^-$, we move from one to another as
if we were attaching the appropriate
handles. In this case ${\Xi}({\hat z})$ behaves
critically as (\ref{Crit}).\par
This criticality stems from the fact that the subleading term
${\lambda}^{(1 - \frac{h[\pi (M^*)]}{2})({\gamma}_0-
2)-1}$,  in the asymptotics of the counting function $B_{\lambda}$,
dominates over the other terms in the statistical sum
when we reach the critical temperature.\par
It is possible to understand this behavior of the grand partition
function on more geometrical grounds by estimating, for $m$
very large, the relative mean-square fluctuation in the filling
function $N_{(m)}$. An elementary computation provides
\begin{eqnarray}
\frac{<({\Delta}N_{(m)})^2>}{(<N_{(m)}>)^2}=
\frac{(\sum_{\lambda}^{\sup N_{(m)}}
{\hat z}^{\lambda}a_{\lambda})(\sum_{\lambda}^{\sup N_{(m)}}
{\hat z}^{\lambda}{\lambda}^2a_{\lambda})}{(\sum_{\lambda}^{\sup
N_{(m)}}
{\hat z}^{\lambda}{\lambda}a_{\lambda})^2}-1
\end{eqnarray}
where $<\ldots>$ denotes the average value with respect to the
statistical sum ${\Xi}(m,\hat{z})$, and
$<({\Delta}N_{(m)})^2> \equiv <N_{(m)}^2>-<N_{(m)}>^2$. At the
critical temperature, and for $\hat{z} \to 1_{-}$, the
leading asymptotic behavior of $<({\Delta}N_{(m)})^2>/(<N_{(m)}>)^2$,
as $m \to \infty$, is provided by, (again by standard
Tauberian theory)
\begin{eqnarray}
\frac{<({\Delta}N_{(m)})^2>}{(<N_{(m)}>)^2}
 = \frac{1}{(1/2)({\gamma}_0 -2)
(2-h[{\pi}_1(M^*)])}
\label{meansquare}
\end{eqnarray}
where $h[{\pi}_1(M^*)] >2$.
Thus, not surprisingly, at the critical regime, the relative
root-mean-square fluctuation in the geodesic balls density is not
negligible, and we encounter unusually large fluctuations in the
filling functions of geodesic balls coverings of manifolds
$d_G$-near $M^*$. According to (\ref{meansquare})
such fluctuations are smaller the larger is
and $h[{\pi}_1(M^*)]$. This does not surprise, since the
larger is the topological
complexity of $M^*$ the less {\it rigid} is the response
of $M^*$ to a deformation towards a nearby, (in the Gromov sense),
topologically
distinct manifold.\par
\vskip 0.5 cm
\section{Connection with 2D-gravity}
In order to discuss the connection between the theory
presented here and two-dimensional quantum gravity models
based on random triangulations of surfaces, let us
examine more explicitly
the physical meaning of the parameters determining the
critical behavior of ${\Xi}({\hat z})$. To this end, it
is convenient to rewrite ${\Xi}$ as
\begin{eqnarray}
 {\Xi}({\hat z})
\sim  {\rho}_h
\Gamma [({\gamma}_0 -2)(1 -\frac{h}{2})]
\left [\frac{({\beta})^{(2-{\gamma}_0)}}{[(c/{\beta}-(1/2)
d(\Omega)]^{({\gamma}_0 -2)}}\right]^{1-
\frac{h[{\pi}_1(M)]}{2}}
\label{roger}
\end{eqnarray}
This rewriting shows that ${\Xi}({\hat z})$, (summed over
the finite number of distinct homotopy types realized by
manifolds in $\Ricco$), formally has the same
structure of the partition function of two-dimensional
quantum gravity if we identify
 $({\beta})^{(2-{\gamma}_0)}$ with a bare
string coupling constant,and  $c/{\beta}$,
(which, up to a minus sign, is the chemical potential for ${\Xi}$),
with a bare cosmological constant.\par
This identification is a consequence of the fact that
through  ${\Xi}({\hat z})$ we are sampling manifold out of
$\Ricco$ according to their fundamental group, and this latter
is probed by the two-skeleton of the geodesic balls coverings.
 {\it As a matter of fact, such two-skeletons are statistically
weighted by
${\Xi}$ through their invariants $N_{(m)}$ and
${\Phi}_{(m)}$ in the same way as in two-dimensional quantum gravity
we weight random  triangulation  of  surfaces}. This
explains the similarity between the two theories. In this
connection, it is also interesting to remark that,
as for two-dimensional gravity,
${\Xi}$ appears to satisfy a scaling relation: ${\Xi}$ depends
on the ratio
\begin{eqnarray}
\frac{({\beta})^{(2-{\gamma}_0)}}{[(c/{\beta}-(1/2)
d(\Omega)]^{({\gamma}_0 -2)}}
\label{scale}
\end{eqnarray}
this remark suggests the question of the existence of a double scaling
limit for ${\Xi}({\hat z})$ analogous to the double scaling limit
in two dimensional quantum gravity which is at the origin of the
progresses in such theory, and of the possible relation between
such kind of limit and the continuum limit dealt with here.\par
 As is known, the double scaling limit in two-dimensional quantum
gravity is taken, as
the cosmological constant approaches its critical value, by letting
the bare string coupling going to zero while keeping fixed a
ratio similar in structure to (\ref{scale}).\par
It is easily checked that such limit corresponds here to having
the ratio (\ref{scale}) finite when, (we keep on in assuming that
$({\gamma}_0-2)<0$), the chemical potential  $c/{\beta}$ is
finite and equal to its critical value $(1/2)d(\Omega)$ and
${\beta} \to \infty$. The geometrical meaning
of this  limiting procedure readily follows from the
expression of the critical inverse
temperature. This latter diverges if
$d_{\lambda} \to d({\Omega}_{(m)})$ as $m \to \infty$,
namely if we sample manifolds in $\Ricco$ with
geodesic balls coverings which are close to being the optimal,
coverings, generated by a collection of close packed geodesic
balls  disposed in a regular array. Roughly speaking, this
implies that the
limit distribution  resulting from this particular
limit procedure samples out of $\Ricco$ manifolds which are
near constant curvature manifolds $M^*$, (the  given
value of the chemical potential fixes the average volume of
the manifolds sampled). Double scaling for ${\Xi}$ thus
appears as a particular subcase, (the zero temperature
limit), of the possible critical regimes that can be
associated with the  large $m$-behavior, on $\Ricco$, of the
statistical
sum ${\Xi}(m,\hat{z})$. \par
\subsection{Some concluding remarks}
Let us  point out some of the problems
we face in trying to correlate the theory described by $\Xi$
to anything which could be reasonably called (euclidean) quantum
gravity. \par
A first problem obviously concerns the choice of Gromov's space
of bounded geometries as the arena for any quantum theory of
gravity. Leaving aside the delicate issue of the Euclidean or
Lorentzian signature, the natural objection which comes about
Gromov's space concerns the restrictions on curvature, diameter, and
volume which appear unnatural in relativity. However, this is not
a real issue, for we can remove such bounds by a limit
procedure. To be more explicit, we can consider a nested sequence
of spaces of bounded geometries
\begin{eqnarray}
\ldots \subset {\cal{R}}(n,r_i,D_i,V_i)\subset {\cal{R}}
(n,r_{i+1},D_{i+1},V_{i+1}) \subset \ldots
\end{eqnarray}
where $r_i \to \infty$, $D_i \to \infty$, and $V_i \to 0$. In
this way we may formally recover the space of all riemannian
structures (completed under Gromov-Hausdorff convergence) as a
limit of spaces of bounded geometries. {\it Manifolds} in such
space may have arbitrarily large curvatures, arbitrarily small and/or
large volumes and diameters, and we can consider the behavior of the
associated sequence of partition functions
${\Xi}_{n,r_i,D_i,V_i}$. The limiting behavior of such sequence
appears quite non-trivial, if , for instance we consider each
${\Xi}_i$ at its critical point, where topology dominates.
Simplyfing a little bit, we can say that the philosophy
underlying the use of the compact Gromov spaces
is simply that such spaces allow for a clear mathematical control
of the {\it cut-off parameters} which determine the behavior, in
the continuum limit, of symplicial approximations to manifolds.\par
\vskip 0.5 cm
A more delicate issue with ${\Xi}$ concerns the {\it regularity}
of the manifolds sampled out in $\Ricco$ according to ${\Xi}$.
More explicitly: when are the (homology) manifolds, dominating
in $\Xi$, {\it smooth} differentiable manifolds?
This problem is already present in two-dimensional quantum gravity,
where
it takes the form of the non-vanishing of the effective string
coupling
constant at the critical point. This yields  for a continuum
limit where the dominating {\it surface} configurations are those
of {\it crumpled} objects where diffeomorphism invariance is
completely
lost. Here, the situation is not very different, in place of crumpled
surfaces we, generally speaking, have to deal with homology
manifolds  (in the physically interesting dimensions three and
four), or with topological manifolds as soon as the dimension equals
five. According to the results of the
previous paragraphs we have some control on what happens at
the critical point to such homology manifolds, but we have no
explicit results which enforce the onset of diffeomorphism
invariance. A rather shy indication in such direction comes
from the geometrical meaning of the double scaling limit which
can be implemented for $\Xi$, but admittedly this is not much.\par

\section*{References}
\begin{description}
\bibitem[1]{Brez}
E.Br\'ezin and V.Kazakov,Phys.\ Lett.\ {\bf B236},144 (1990).
\bibitem[2]{Doug}
M.Douglas and S.Shenker, Nucl.\ Phys.\ {\bf B235}, 635 (1990).
\bibitem[3]{Gros}
D.J.Gross and A.A.Migdal, Phys.\ Rev.\ Lett. {\bf 64}, 127 (1990);
Nucl.\ Phys.\  {\bf B340}, 333 (1990).
\bibitem[4]{Carf}
M.Carfora and A.Marzuoli, {\it Bounded geometries and topology
fluctuations in lattice quantum gravity} to appear in
Class.Quant. Grav. (1991); see also
 {\it Critical phenomena for Riemannian
geometries} Preprint FNT/T 90/07, (Univ. Pavia). \par
A more detailed mathematical presentation can be found in \par
M.Carfora and A.Marzuoli, {\it Finiteness theorems in
Riemannian geometry and lattice quantum gravity}, to appear in
Contemporary Mathematics (Proceedings of the AMS research meeting
{\it Mathematical Aspects of Classical Field Theory}, Seattle, 1991).
\bibitem[5]{Marz}
M.Carfora and A.Marzuoli, Phys.\ Rev.\ Lett.\  {\bf 62}, 1339 (1989).
\bibitem[6]{Grom}
M.Gromov, {\it Structures m\'etriques pour les vari\'et\'es
Riemanniennes} (Conception Edition Diffusion Information
Communication Nathan, Paris, 1981).\par
See also
S.Gallot, D.Hulin, J.Lafontaine {\it Riemannian Geometry},
(Springer Verlag, New York,1987).\par
For the convenience of the
reader we recall that the diameter of a riemannian manifold $M$ is
defined as $\sup_{(p,q)\in M\times M}d(p,q)$ where $d(\cdot,\cdot)$
denotes the distance function of $M$.
\bibitem[7]{Grov}
K.Grove and P.V.Petersen, Annals of Math.\  {\bf 128}, 195 (1988).
\bibitem[8] {Erdo}
P.Erdos, P.Kelly, Amer.Math.Monthly {\bf 70} 1074 (1963);
see also chap.10 of F.Harary (ed.) {\it A Seminar on Graph
Theory} (Holt, Rinehart and Winston, Inc.,New York, 1967).
\bibitem[9]{Ruel}
D.Ruelle,{\it Statistical Mechanics: rigorous results}
(W.A.Benjamin, New York, 1974).
\bibitem[10]{Pari}
D.Bessin, C.Itzykson and J.B.Zuber, Adv.\ Appl.\ Math.\
 {\bf 1},109(1980).
\bibitem[11]{Hara}
F.Harary and E.Palmer, {\it Graphical Enumeration} (Academic, New
York, 1973).
\bibitem[12]{Pete}
K.Grove, P.V.Petersen and J.Y.Wu, Bull.\ Am.\ Math.\ Soc.\  {\bf 20},
181
(1989).
\bibitem[13] {Ambj}
J.Ambjorn, {\it Random surfaces: a non-perturbative
regularization of strings} in {\it Probabilistic Methods in
Quantum Field Theory and Quantum Gravity} P.H.Damgaard {\it et al.}
eds. (Plenum Press, New York, 1990).
\bibitem[14] {Pete}
K.Grove, P.V.Petersen and J.Y.Wu, Bull.Am.Math.Soc. {\bf 20}, 181
(1989); Invent.Math. {\bf 99}, 205 (1990) (and its Erratum).
\bibitem[15]{Rema}
It is interesting to note
that the structure of
$\Ricco$ is quite sensible to the dimension $n$, in the sense that
 for $n \geq 5$ one can  actually count (at least
in principle), the different types of topologies and the different
types of differentiable structures realized in each homotopy
class of manifolds sampled by ${\Xi}(m,z)$, since the number of
such structures can be shown to be finite. In dimension
three and four this counting is no longer
possible.Some remarks concerning the phase structure of high-
dimensional,
($n \geq 5)$, lattice quantum gravity theories will be discussed
elsewhere.
\end{description}
\end{document}